\pdfoutput=1
\RequirePackage[T1]{fontenc}
\RequirePackage{fix-cm}
\RequirePackage{lmodern}
\documentclass{ieeeaccess}
\usepackage{cite}
\usepackage{amsmath,amssymb,amsfonts}
\usepackage{graphicx}
\usepackage{textcomp}
\def\BibTeX{{\rm B\kern-.05em{\sc i\kern-.025em b}\kern-.08em
    T\kern-.1667em\lower.7ex\hbox{E}\kern-.125emX}}
\begin{document}
\history{Received: July 31, 2024; Revised: September 19, November 22, and December 17, 2024, Accepted: January 13, 2025.}
\doi{10.1109/TQE.2025.DOI}

\newcommand{\ket}[1]{|#1 \rangle}
\newcommand{\bra}[1]{\langle #1 |}
\newtheorem{theorem}{Theorem}
\newtheorem{proposition}[theorem]{Proposition}
\newtheorem{remark}[theorem]{Remark}
\newtheorem{lemma}[theorem]{Lemma}

\title{Advance Sharing Procedures for the Ramp Quantum Secret Sharing Schemes
With the Highest Coding Rate}
\author{\uppercase{Ryutaroh Matsumoto}\authorrefmark{1}, \IEEEmembership{Member, IEEE}}
\address[1]{Department of Information and Communications Engineering,
  Institue of Science Tokyo, Tokyo 152-8550 Japan (email: ryutaroh@ict.e.titech.ac.jp)}
\tfootnote{This work was supported in part by the JSPS grant No.\ 23K10980.}

\markboth
{R.\ Matsumoto \headeretal: Advance Sharing of Quantum Shares}
{R.\ Matsumoto \headeretal: Advance Sharing of Quantum Shares}

\corresp{Corresponding author: Ryutaroh Matsumoto (email: ryutaroh@ict.e.titech.ac.jp).}

\begin{abstract}
In some quantum secret sharing schemes,
  it is known that some shares can be distributed to participants before
  a secret is given to the dealer.
  However, it is unclear whether some shares can be distributed before
  a secret is given in the ramp quantum secret sharing schemes
  with the highest coding rate.
  In this paper, we propose procedures to distribute some shares
  before
  a secret is given in those schemes.
    The new  procedures enhances applicability of the secret sharing schemes
    to wider scenarios as some participants can be unavailable when the dealer obtains
    the quantum secret.
  Then we prove that our new encoding procedures retain the correspondences
  between quantum secrets and quantum shares in the original schemes,
  which ensures the highest coding rates of the original schemes are also retained.
\end{abstract}


\titlepgskip=-15pt

\maketitle

\section{Introduction}\label{sec1}
The study of quantum information processing gains
much attention recently.
One reason is the increase in the size of quantum computers
\cite{googleqc}.
The storage and communications of quantum information
are still difficult experimentally, but they will probably become easier and less expensive
in a future.
Classical secret sharing, in which secrets and shares are both classical information,
is nowadays used in practice, for example, in distributed storage systems.
In a distributed storage system \cite{attasena2017}, data are stored in multiple storages,
which increases the risk of data leakage. Secret sharing schemes decreases that risk
with the multiple storages.
It is likely that quantum secret sharing schemes play similar roles
in the future quantum information era.

\begin{table*}
  \caption{Comparisons of the original and the proposed encoding procedures}\label{tab1}
  \begin{tabular}{rcccc}
    &Ogawa et al.\ \cite{OGAWA05}&Zhang-Matsumoto \cite{MATSUMOTO14STRONG}&
    Section \ref{sec2}&Section \ref{sec3}\\\hline
    dimension of a quantum share &$q$&$q$&$q$&$q$\\
    number of symbols in quantum secrets&$L$&$L$&$L$&$L$\\
    maximum number of participants&$q-1$&$q-L$&$q-1$&$q-L$\\
    coding rate&$L$&$L$&$L$&$L$\\
    minimum number of participants to reconstruct secrets&$k$&$k$&$k$&$k$\\
    maximum number of participants with no information about secrets&$k-L$&$k-L$&$k-L$&$k-L$\\
    strong security&No&Yes&No&Yes\\
    maximum number of shares distributed in advance &$0$&$0$&$k-L$&$k-L$\\\hline
  \end{tabular}
  \end{table*}

In this paper, we consider to share a quantum secret by quantum shares.
We assume no interactions among the dealer and the participants
except distribution of quantum shares from the dealer to the participants.
Such a problem formulation was initiated by Cleve, Gottesman and Lo
\cite{cleve99,gottesman00}, which is the most natural quantum
counterpart of the classical secret sharing
considered by Shamir \cite{shamir79} and Blakley \cite{blakley79}.

The coding rate is an important parameter in secret sharing schemes.
It is defined as the ratio of secret size to average of share sizes \cite{cleve99,gottesman00,OGAWA05}. Higher coding rates are desirable.
Another important property of secret sharing schemes
is the access structure, which consists of three families of qualified sets, forbidden sets,
and intermediate sets \cite{OGAWA05}.
A set $S$ of shares is called qualified (resp.\  forbidden)
if $S$ allows reconstruction of the secret (resp.\ has no information about the secret).
A set $S$ of shares is called intermediate if it is neither qualified nor forbidden.
If a secret sharing scheme has no intermediate set, it is called perfect.
The coding rate cannot be greater than $1$ if it is perfect \cite{gottesman00}.
Ogawa et al.\ proposed ramp quantum secret sharing schemes,
which enable coding rates greater than $1$, at the cost of allowing intermediate sets.
In this paper, we consider $q$-dimensional quantum system $\mathcal{H}_q$
and call it a qudit, where $q$ is a prime power.
A $(k, L, n)$ ramp quantum secret sharing scheme encodes
$L$-qudit quantum secret into $n$ shares each of which is 1 qudit in $\mathcal{H}_q$.
The parameter $k$ decides the access structure mentioned earlier,
as follows.
In a $(k, L, n)$ scheme, a set $S$ is qualified if $|S| \geq k$, forbidden if $|S| \leq k-L$,
and intermediate otherwise (i.e.\ $k-L+1 \leq |S| \leq k-1$).
Among $(k, L, n)$ schemes, Ogawa et al.'s one has the highest possible coding
rate $L$, and firstly we consider Ogawa et al.'s ramp quantum secret sharing
in Section \ref{sec2}.
The scheme by Cleve, Gottesman and Lo \cite{cleve99} is a special
case of Ogawa et al.'s corresponding to the case $L=1$.

As mentioned before,
an intermediate set  has nonzero information about the secret in a ramp scheme.
In classical secret sharing scheme, when the secret $\vec{s} = (s_1$, \ldots, $s_L)$,
Iwamoto and Yamamoto \cite{iwamoto06} showed an explicit example of ramp schemes
in which a component $s_i$ in the secret can be reconstructed from
an intermediate set $S$.
In order to prevent such information leakage, Yamamoto \cite{yamamoto86}
defined the strong security for $(k,L,n)$ classical ramp secret sharing scheme,
in which any set $S$ of shares has no information about part
$(s_i : i \in T)$ of the secret if $|S| + |T| \leq k$.
In the context of ramp quantum secret sharing,
Zhang and Matsumoto \cite{MATSUMOTO14STRONG}
showed an explicit example from Ogawa et al.'s scheme in which an intermediate
set leaks part of the quantum secret similarly to the classical case in this paragraph.
They also introduced a strong security definition into ramp quantum secret sharing,
and explicitly construct a $(k,L,n)$ strongly secure ramp quantum secret sharing
scheme whose coding rate $L$ is as high as Ogawa et al.'s scheme \cite{OGAWA05}.
On the other hand, the Zhang-Matsumoto scheme has a more stringent
condition on the number $n$ of participants, that is, $n\leq q-L$,
while Ogawa et al.'s scheme has a milder condition $n \leq q-1$.
The condition $n \leq q-1$ also exists in Shamir's scheme
\cite{shamir79,blakley79}, but in the classical case $q$ can be made almost
arbitrarily large.
On the other hand,
because the quantum system $\mathcal{H}_q$ cannot be chosen freely,
the symbol size $q$ cannot be adjusted in the quantum case
as easily as the classical case.
Therefore, both Ogawa et al.'s scheme and Zhang-Matsumoto scheme are
practically useful depending on applications.

It is sometimes convenient to distribute shares before a secret is given
to the dealer. One example of such situations was discussed in
\cite{miyajima22}, which considered sharing a classical secret
by quantum shares. Such distribution of shares before a given secret 
is named ``advance sharing'' \cite{miyajima22}.
Advance sharing is a trivial problem when both secret and shares are classical.
For example, encoding 1-bit secret $s$ into 2 shares $(x, s+x)$ provides a $(1,2)$
perfect secret sharing scheme and the first share $x$ is clearly advance shareable,
where $x$ is a random bit.
Advance sharing enhances applicability of secret sharing schemes to
wider scenarios.

On the other hand, when secrets are quantum, it is nontrivial to
realize advance sharing. In this paper, we focus on the case of quantum secrets.
The first quantum advance sharing was realized in Lie and Jeong
\cite{lie20}, which enabled advance sharing for some specific schemes.
On the other hand, any quantum error-correcting code and
its erasure decoding algorithm can be used as a quantum secret sharing
scheme \cite{cleve99,gottesman00}, and Shibata and Matsumoto \cite{shibata24}
clarified how to realize advance sharing with quantum secret sharing scheme
constructed from a $p$-adic quantum stabilizer code,
where $p$ is a prime number.
Based on \cite{shibata24}, Masumori and Matsumoto  \cite{isita2024}
showed how to realize advance sharing with a limited special case of
Ogawa et al.\ \cite{OGAWA05}, where the dimension $q$ of $\mathcal{H}_q$
was restricted to  a prime number.
Because the general method \cite{shibata24} for advance sharing
cannot handle stabilizer codes over $\mathcal{H}_q$ with non-prime $q$,
the previous proposal \cite{isita2024} of advance sharing for Ogawa et al.\ \cite{OGAWA05}
cannot be immediately extended to the general case of Ogawa et al.\ \cite{OGAWA05}.

When we consider advance sharing for the strongly secure scheme
\cite{MATSUMOTO14STRONG}, another limitation of \cite{shibata24}
appears. The strong security property in \cite{MATSUMOTO14STRONG}
is realized by a careful correspondence between quantum secrets
and quantum shares. On the other hand,
the general method \cite{shibata24} for advance sharing
destroys the correspondence between quantum secrets
and quantum shares when it is applied to a ramp quantum secret sharing
scheme. The general method \cite{shibata24} seems almost impossible
to be applied
for realizing advance sharing with the strongly secure scheme \cite{MATSUMOTO14STRONG}.

In this paper,
we propose new encoding procedures of quantum secrets
into quantum shares for realizing advance sharing with
Ogawa et al.'s scheme \cite{OGAWA05} and Zhang-Matsumoto scheme
\cite{MATSUMOTO14STRONG}.
The new procedures enhance applicability of the scheme proposed in \cite{OGAWA05,MATSUMOTO14STRONG}
to wider scenarios in which some participants are unavailable when the dealer obtains quantum secrets
to be shared.
Both procedures retain the correspondence between quantum secrets
and quantum shares in the original schemes.
Therefore, all properties in the original schemes, such as coding rates,
access structures, strong security, etc., remain the same as the originals.
In particular, the highest coding rates are retained from the original schemes
\cite{OGAWA05,MATSUMOTO14STRONG}.
The proposed new encoding procedures add
extra useful functionalities to the original ramp quantum schemes
\cite{OGAWA05,MATSUMOTO14STRONG}.
The differences among the original schemes \cite{OGAWA05,MATSUMOTO14STRONG}
and the proposals is summarized in Table \ref{tab1}.

This paper is organized as follows:
In Section \ref{sec2}, we review necessary contents from
Ogawa et al.\ \cite{OGAWA05}, a new encoding procedure is given for \cite{OGAWA05},
and  we prove that the correspondence between secrets and shares
is retained from \cite{OGAWA05}.
Section \ref{sec3} has a similar structure to Section \ref{sec2}.
In Section \ref{sec3}, we review \cite{MATSUMOTO14STRONG},
propose a new encoding procedure for \cite{MATSUMOTO14STRONG},
then we prove that the same correspondence between secrets and shares.
Throughout Sections \ref{sec2} and \ref{sec3},
illustrating examples of the original and the proposed encoding are included.
In Section \ref{sec4}, we give concluding remarks and future research agendas.

\section{Advance sharing with quantum ramp secret sharing scheme in \protect\cite{OGAWA05}}\label{sec2}
\subsection{Short review of quantum secret sharing}\label{sec21}
In this subsection,
we briefly describe the quantum secret sharing
considered in this paper, and its security model of secret sharing schemes.
In short,
it is exactly the same as the paper by Cleve, Gottesman, and Lo
\cite{cleve99} and Ogawa et al.\ \cite{OGAWA05}.
When both secret and shares are classical information,
this model is equivalent to Shamir-Blakley scheme
\cite{stinson06}.

We always assume that there are $n$ participants and each
participant receives one share
from the dealer.
The entire set of shares/participants
are denoted by $\{1$, \ldots, $n\}$.
Therefore, any subset $A \subset \{1$, \ldots, $n\}$ corresponds to
some set of shares/participants.

Let $q$ be a prime power, $\mathbf{F}_q$ be the finite field with $q$ elements,
and $\{ \ket{i} : i \in \mathbf{F}_q \}$ be an orthonormal basis
of $\mathcal{H}_q$.
For $\vec{s}=(s_1$, \ldots, $s_L) \in \mathbf{F}_q^L$
we define
\[
\ket{\vec{s}} = \ket{s_1} \otimes \cdots \otimes \ket{s_L} \in
\mathcal{H}_q^{\otimes L}.
\]
The dealer encodes the quantum secret $\ket{\vec{s}}$
(or a linear combination of $\ket{\vec{s}}$)
into some pure state $\ket{\varphi} \in \mathcal{H}_q^{\otimes n}$.
The dealer distributes each qudit in $\ket{\varphi}$ to each participant.
Observe that no measurement, classical communications,
nor teleportation of quantum states are involved here,
in contrast to \cite{hillery99}.
Also observe that the purpose and the procedure are different from
those in quantum secure direct communication (QSDC) \cite{qsdcsurvey2024}.
There seems no explicit relationship between
the QSDC and the quantum secret sharing considered here.

A set $A \subset \{1$, \ldots, $n\}$ is said to be qualified
if $\ket{\vec{s}}$ can be reconstructed from
$\mathrm{Tr}_{\overline{A}}[\ket{\varphi}\bra{\varphi}]$,
said to be forbidden if
$\mathrm{Tr}_{\overline{A}}[\ket{\varphi}\bra{\varphi}] $ is independent of
$\ket{\vec{s}}$, and said to be intermediate if $A$ is neither qualified nor
forbidden \cite{OGAWA05},
where $\mathrm{Tr}_{\overline{A}}$ denotes the partial trace
over $\overline{A} = \{1$, \ldots, $n\} \setminus A$.
As mentioned in Section \ref{sec1},
a quantum secret sharing scheme is said to be a ramp scheme
if there is no intermediate set.

In Sections \ref{sec25} and \ref{sec35},
we will show that our proposed encoding procedures retain the correspondences
between quantum secrets and quantum shares from
\cite{OGAWA05,MATSUMOTO14STRONG}.
This means that the sets of qualified sets and forbidden sets remains the same
as \cite{OGAWA05,MATSUMOTO14STRONG}.

More importantly, since the correspondences between secrets and shares
are the same as \cite{OGAWA05,MATSUMOTO14STRONG},
new encoding procedures are as much secure as
the original ones under any security models and assumptions,
and any new security argument is unnecessary unless a security property
not considered in  \cite{OGAWA05,MATSUMOTO14STRONG}
is required.

Although we can consider cheating participants among
legitimate ones
and adversaries outside
of legitimate participants \cite{crepeau02,benor06},
we do not consider these scenarios.
We do not consider approximate quantum secret sharing
\cite{crepeau05}, neither,
as they are not considered in the original encoding procedures
\cite{OGAWA05,MATSUMOTO14STRONG}.
However, since those other scenarios are also important,
advance sharing in them should also be investigated in future.

The standard metrics for evaluating secret sharing schemes are
coding rates and access structures \cite{OGAWA05},
and sometimes the strong security \cite{yamamoto86,MATSUMOTO14STRONG}.
The proposed encoding procedures retains those evaluation metrics from
the original schemes \cite{OGAWA05,MATSUMOTO14STRONG},
and they enable advance sharing that enhances applicability of secret sharing schemes
to wider scenarios.

\subsection{Ogawa et al.'s scheme}
Let $n$ be the number of shares/participants, and we assume $n\leq q-1$.
Let $\alpha_1$, \ldots, $\alpha_n$ be distinct nonzero\footnote{%
Ogawa et al.\ \cite{OGAWA05} allowed zero but it was a mistake.}
elements in $\mathbf{F}_q$.
We will review the construction of a $(k,L,n)$ ramp quantum scheme.
For $\vec{c} = (c_1$, \ldots, $c_k) \in \mathbf{F}_q^k$,
we define the polynomial $f_{\vec{c}}(x)$ by
\begin{equation}
f_{\vec{c}}(x) = c_1 + c_2x+ \cdots + c_k x^{k-1}.\label{eq9}
\end{equation}
A quantum secret $\ket{\vec{s}} \in \mathcal{H}_q^{\otimes L}$
is encoded to
\begin{equation}
  \frac{1}{\sqrt{q^{k-L}}}\sum_{\vec{c}\in D(\vec{s})} \ket{f_{\vec{c}}(\alpha_1)} \otimes \ket{f_{\vec{c}}(\alpha_2)} \otimes \cdots \otimes \ket{f_{\vec{c}}(\alpha_n)}, \label{eq10}
\end{equation}
where $D(\vec{s})$ is the set of vectors $\vec{c} \in \mathbf{F}_q^k$
whose leftmost $L$ components are the same as those in $\vec{s}$.
The quantum state in Eq.\  (\ref{eq10}) consists of $n$ qudits.
The $i$-th qudit in Eq.\  (\ref{eq10}) is distributed to the $i$-th participant.

\subsection{Example of the original encoding}\label{ex:original}
Let $q=4$, $n=3$, $k=2$, $L=1$, and
$\alpha_i = \alpha^i \in \mathbf{F}_4$,
where $\alpha$ is a primitive element of $\mathbf{F}_4$.
The standard encoding procedure in Eq.\ (\ref{eq10})
encodes a quantum secret $\ket{\vec{s}}$ for $\vec{s} =(s) \in \mathbf{F}_4^1$
into
$\frac{1}{\sqrt{4}}\sum_{\vec{c}\in D(\vec{s})} \ket{f_{\vec{c}}(\alpha_1)} \otimes \ket{f_{\vec{c}}(\alpha_2)} \otimes \ket{f_{\vec{c}}(\alpha_3)} \nonumber =
    \frac{1}{\sqrt{4}}(
    \ket{s$, $ s$, $ s} +
    \ket{s+\alpha$, $ s+\alpha^2$, $ s+1} +
    \ket{s+\alpha^2$, $ s+1$, $ s+\alpha} +
    \ket{s+1$, $ s+\alpha$, $ s+\alpha^2}). $
As every qudit  seems to depend on the quantum secret $\ket{\vec{s}}$,
from the encoding procedure in Ogawa et al.\ \cite{OGAWA05}, it seems unclear
how one share is advance shareable.

\subsection{Proposed advance sharing procedure for \protect\cite{OGAWA05}}
We will propose an advance sharing procedure which retains
the correspondence between a quantum secret and quantum shares
given in Eq.\ (\ref{eq10}),
where the $(k,L,n)$ quantum ramp scheme was considered.

We will introduce an elementary lemma in linear algebra and polynomials.
  \begin{lemma}\label{lem11}
    Let $\mathcal{P}_m$ be the set of univariate polynomials $f(x)$
    over $\mathbf{F}_q$ with $\deg(f) < m$. Consider the evaluation map
    $\mathrm{ev}(f) = (f(\alpha_1)$, \ldots, $f(\alpha_n))$,
    where $\alpha_1$, \ldots, $\alpha_n$ are pairwise distinct as before
    while $\alpha_1$, \ldots, $\alpha_n$ may contain $0 \in \mathbf{F}_q$.
    The map $\mathrm{ev}$ is a linear map from $\mathcal{P}_m$
    to $\mathbf{F}_q^n$.
    If $m\leq n$ then $\mathrm{ev}$ is injective,
    and if $m\geq n$ then $\mathrm{ev}$ is surjective.
  \end{lemma}
  \begin{IEEEproof}
    Suppose that $m\leq n$. If $\mathrm{ev}(f) = \vec{0}$
    then $f(\alpha_1) = \cdots = f(\alpha_n)=0$.
    Since $\deg(f) < m \leq n$, this means that $f(x)$ is the zero polynomial.
    We have seen that $\ker(\mathrm{ev}) = \{0\}$ when $m\leq n$,
    which means that $\mathrm{ev}$ is an injective linear map.

    By the previous discussion, $\mathrm{ev}$ is injective
    on $\mathcal{P}_n$. Since $\dim \mathcal{P}_n = n$,
    the image $\mathrm{ev}(\mathcal{P}_n)$ of $\mathcal{P}_n$
    under the map $\mathrm{ev}$ is $\mathbf{F}_q^n$.
    This means that $\mathrm{ev}$ is surjective on $\mathcal{P}_m$ if
    $m\geq n$.
  \end{IEEEproof}

We consider to distribute $k-L$ shares before $\ket{\vec{s}}$ is given to the
dealer. By reassigning indices, we may assume that the dealer
wants to advance share the 1st to the $(k-L)$-th shares.
Let
\begin{equation}
\ket{\Psi} = \frac{1}{\sqrt{q^{k-L}}}
\sum_{\vec{r}\in \mathbf{F}_q^{k-L}} \ket{\vec{r}} \otimes \ket{\vec{r}}, \label{eq31}
\end{equation}
where $\ket{\vec{r}} = \ket{r_1} \otimes \cdots \otimes \ket{r_{k-L}}$
for $\vec{r} = (r_1$, \ldots, $r_{k-L})\in \mathbf{F}_q^{k-L}$.
Note that  $\ket{\Psi}$ consists of $2(k-L)$ qudits, as $\ket{\vec{r}}$
consists of $(k-L)$ qudits.

Suppose that the quantum secret is
$\ket{\vec{s}} = \ket{s_1} \otimes \cdots \otimes \ket{s_L} \in
\mathcal{H}_q^{\otimes L}$
with $\vec{s}=(s_1$, \ldots, $s_L) \in \mathbf{F}_q^L$.
For $\vec{r}\in \mathbf{F}_q^{k-L}$
and $\vec{s}$, let $g_{\vec{r},\vec{s}}(x)$ be a  univariate polynomial
such that $g_{\vec{r},\vec{s}}(x) \in \{ f_{\vec{c}}(x) : \vec{c} \in D(\vec{s})\}$ and
$g_{\vec{r},\vec{s}}(x)(\alpha_i) = r_i$
  for $i=1$, \ldots, $k-L$.
  
  \begin{proposition}\label{lem10}
    For given $\vec{r}$ and $\vec{s}$,
    the polynomial $g_{\vec{r},\vec{s}}(x)$ exists uniquely.
  \end{proposition}
  \begin{IEEEproof}
    Let $f_{\vec{c}}(x)$ for $\vec{c}\in D(\vec{s})$ as defined in Eqs.\ (\ref{eq9}) and
    (\ref{eq10}).
    Recall that the coefficients $c_1$, \ldots, $c_L$ are known
    as $c_1=s_1$, \ldots, $c_L=s_L$ by the
    condition $\vec{c} \in D(\vec{s})$.
    We will regard the coefficients $c_{L+1}$, \ldots, $c_k$ in $f_{\vec{c}}(x)$
    as $k-L$ unknowns in a system of linear equations.
    For $i=1$, \ldots, $k-L$,
    the condition $g_{\vec{r},\vec{s}}(x)(\alpha_i) = r_i$
    can be written as
    \begin{eqnarray}
    &&c_{L+1} + c_{L+1}\alpha_i + \cdots + c_k \alpha_i^{k-L-1}\\
    &=&  \frac{r_i - s_1 - s_2 \alpha_i - \cdots - s_L \alpha_i^{L-1}}{\alpha_i^L}.
    \label{eq1000}
    \end{eqnarray}
      The polynomial $f(x) = c_{L+1} + \cdots + c_k x^{k-L-1} \in \mathcal{P}_{k-L}$
      corresponds to a solution of the system of linear equations
      if $\mathrm{ev}(f) =
      ((r_1-s_1 - s_2 \alpha_1 - \cdots - s_L \alpha_1^{L-1})/\alpha_1^L$,
      $ (r_2 - s_1 - s_2 \alpha_2 - \cdots - s_L \alpha_2^{L-1})/\alpha_2^L$, \ldots,
      $(r_{k-L} - s_1 - s_2 \alpha_{k-L} - \cdots - s_L \alpha_{k-L}^{L-1})/\alpha_{k-L}^L)$,
      where the map $\mathrm{ev}$ is from $\mathcal{P}_{k-L}$ to
      $\mathbf{F}_q^{k-L}$. By Lemma \ref{lem11}
      we can see that the solution of the system of linear equations considered here
      exists uniquely,
    which ensures unique existence of $g_{\vec{r},\vec{s}}(x)$
    for given $\vec{r}$ and $\vec{s}$.
    \end{IEEEproof}

\begin{remark}\label{rem10}
The coefficients of $g_{\vec{r},\vec{s}}(x)$
can be more explicitly described in terms of $\vec{r}$ and $\vec{s}$,
as follows.
As  in the proof of Proposition \ref{lem10},
let $f(x) = c_{L+1} + \cdots + c_k x^{k-L-1}$.
We also use
$(b_1$, \ldots, $b_{k-L}) = ((r_1-s_1 - s_2 \alpha_1 - \cdots - s_L \alpha_1^{L-1})/\alpha_1^L$,
      $ (r_2 - s_1 - s_2 \alpha_2 - \cdots - s_L \alpha_2^{L-1})/\alpha_2^L$, \ldots,
      $(r_{k-L} - s_1 - s_2 \alpha_{k-L} - \cdots - s_L \alpha_{k-L}^{L-1})/\alpha_{k-L}^L)$.
As shown in the proof of Proposition \ref{lem10},
we have $f(\alpha_i) = b_i$ for $i=1$, \ldots, $k-L$.
To determine $f(x)$, we can use the Lagrange interpolation formula
\cite[Chapter 13]{stinson06}.
For $i=1, \ldots, k-L$,
let
\[
\ell_i(x) = \prod_{1 \leq j \leq k-L, j\neq i} \frac{x-\alpha_j}{\alpha_i-\alpha_j}.
\]
Then $f(x) = c_{L+1} + \cdots + c_k x^{k-L-1}$
is given as
\begin{eqnarray*}
f(x)&=& \sum_{i=1}^{k-L}b_i \ell_i(x)\\
&=& \sum_{i=1}^{k-L}\frac{r_{i} - s_1 - s_2 \alpha_{i} - \cdots - s_L \alpha_{i}^{L-1}}{\alpha_{i}^L} \ell_i(x)
\end{eqnarray*}
and  we have
\begin{eqnarray*}
&&g_{\vec{r},\vec{s}}(x)\\
&=&  s_1 + s_2 x + \cdots +
s_L x^{L-1} + c_{L+1} x^L + \cdots + c_k x^{k-1}\\
&=& x^L f(x) + \sum_{i=1}^L s_i x^{i-1}.
\end{eqnarray*}
\end{remark}

Now we are ready to describe a new encoding procedure enabling
advance sharing.
    By Proposition \ref{lem10},
    we can define a unitary map $U_{\mathrm{enc}}$ from
    $\mathcal{H}_q^{\otimes k}$ to  $\mathcal{H}_q^{\otimes (n-k+L)}$ 
    sending a quantum state $\ket{\vec{r}}\ket{\vec{s}}$
    to $\ket{g_{\vec{r},\vec{s}}(\alpha_{k-L+1})} \otimes \cdots \otimes
    \ket{g_{\vec{r},\vec{s}}(\alpha_n)}$.
    Note that $k=n-k+L$ by \cite[Lemma 2]{OGAWA05}.
  
The proposed procedure for advance sharing is as follows:
\begin{enumerate}
\item The dealer distributes $k-L$ qudits in the left half of Eq.\ (\ref{eq31}).
\item After the quantum secret $\ket{\vec{s}}$ is given,
the dealer applies $U_{\mathrm{enc}}$ to the remaining half of Eq.\ (\ref{eq31})
and $\ket{\vec{s}}$. Then the quantum state of all the $n$ shares becomes
  \begin{equation}
  \frac{1}{\sqrt{q^{k-L}}}
  \sum_{\vec{r}\in \mathbf{F}_q^{k-L}} \ket{\vec{r}}\otimes \ket{g_{\vec{r},\vec{s}}(\alpha_{k-L+1})}
    \otimes \cdots \otimes \ket{g_{\vec{r},\vec{s}}(\alpha_{n})}. \label{eq33}
  \end{equation}
\end{enumerate}

\begin{remark}\label{rem1}
When $k-L-1$ or a fewer shares are advance shared,
the dealer can simply keep some shares in the advance sharing phase in our
proposal.
On the other hand, it is impossible to advance share $k-L+1$ or more shares.
When a set $J$ of shares is not forbidden, the shares in $J$ depend on quantum secrets
\cite[Theorem 2]{OGAWA05}.
Thus, in order for a set $J$ of shares to be advance shareable, $J$ must be a forbidden set.
In Ogawa et~al.'s $(k,L,n)$ scheme, any $k-L+1$ or more shares
cannot form a forbidden set,
and cannot be advance shareable.
Our proposal makes advance shareable sets as large as possible.
\end{remark}

\subsection{Proof of correctness}\label{sec25}
It is clear from the last subsection that the proposed procedure can distribute
$k-L$ or a fewer shares before a secret is given to the dealer.
On the other hand, at this point it is unknown whether or not
the proposed encoding procedure produces the same quantum states
of shares as the original procedure \cite{OGAWA05}.
In order to show their sameness,
in this subsection, we will prove that
the proposed procedure gives the same quantum state of $n$ shares
as the original scheme by Ogawa et al.\ \cite{OGAWA05}
for a given quantum secret $\ket{\vec{s}}$.
In Eq.\ (\ref{eq10}), the set of indices appearing in the quantum state is
\begin{equation}
V_1 = \{(f_{\vec{c}}(\alpha_1), \ldots, f_{\vec{c}}(\alpha_n) ) : \vec{c}\in D(\vec{s})\}.
\label{eq41}
\end{equation}
  In Eq.\ (\ref{eq33}), the set of indices appearing in the quantum state is
\begin{eqnarray}
V_2&=& \{(r_1, \ldots, r_{k-L},   g_{\vec{r},\vec{s}}(\alpha_{k-L+1}), \ldots, g_{\vec{r},\vec{s}}(\alpha_{n}))\nonumber\\
&&  : \vec{r} \in \mathbf{F}_q^{k-L} \}. \label{eq42}
\end{eqnarray}
\begin{theorem}
For a fixed quantum secret $\ket{\vec{s}}$ for $\vec{s} \in \mathbf{F}_q^L$,
the proposed procedure gives the same quantum state of shares
as the original scheme by Ogawa et al.\ \cite{OGAWA05}.
\end{theorem}
\begin{IEEEproof}
Equation  (\ref{eq10}) can be written as
\begin{equation}
\frac{1}{\sqrt{q^{k-L}}}
\sum_{(v_1, \ldots, v_n) \in V_1} \ket{v_1} \otimes \cdots \otimes \ket{v_n},\label{eq1001}
\end{equation}
and Eq.\ (\ref{eq33})
 can be written as
\begin{equation}
\frac{1}{\sqrt{q^{k-L}}}
\sum_{(v_1, \ldots, v_n) \in V_2} \ket{v_1} \otimes \cdots \otimes \ket{v_n}.\label{eq1002}
\end{equation}
To prove the theorem, we have to prove that
Eqs.\ (\ref{eq1001}) and (\ref{eq1002}) are equal.
In order to prove the equality between Eqs.\ (\ref{eq1001}) and (\ref{eq1002}),
it is enough to show that the sets (\ref{eq41}) and (\ref{eq42}) are the same.
To show the equality between two sets (\ref{eq41}) and (\ref{eq42}),
in Eq. (\ref{eq1003}) we will show that there is a one-to-one correspondence
between elements in the two sets (\ref{eq41}) and (\ref{eq42}).

For fixed $\vec{s} \in \mathbf{F}_q^L$ and
$\vec{r} \in \mathbf{F}_q^{k-L}$, by Proposition \ref{lem10}
there exists a unique polynomial $f_{\vec{c}}$ with $\vec{c} \in D(\vec{s}) $.
This correspondence gives a bijection  sending $\vec{r} \in \mathbf{F}_q^{k-L}$ to
$f_{\vec{c}}$ with $\vec{c} \in D(\vec{s}) $.
By the definition of $g_{\vec{r},\vec{s}}$, we have
\begin{eqnarray}
&&(r_1, \ldots, r_{k-L},   g_{\vec{r},\vec{s}}(\alpha_{k-L+1}), \ldots, g_{\vec{r},\vec{s}}(\alpha_{n})) \in V_2 \nonumber\\
&=& (g_{\vec{r},\vec{s}}(\alpha_1), \ldots, g_{\vec{r},\vec{s}}(\alpha_{n}))\nonumber\\
&=&(f_{\vec{c}}(\alpha_1), \ldots, f_{\vec{c}}(\alpha_{n})) \in V_1
\label{eq1003},
\end{eqnarray}
which shows the theorem, where $\vec{c}$ is chosen according to $\vec{r}$.
\end{IEEEproof}

\subsection{Example of the proposed encoding}
We reuse definitions from Section \ref{ex:original}.
Let $q=4$, $n=3$, $k=2$, $L=1$, and
$\alpha_i = \alpha^i \in \mathbf{F}_4$,
where $\alpha$ is a primitive element of $\mathbf{F}_4$.
Note that since $q$ is not a prime number, this case cannot be handled in
the previous research \cite{isita2024}.
In our proposal,
the dealer prepares 2-qudit entangled state
$\frac{1}{2}\sum_{r \in \mathbf{F}_4} \ket{r}\ket{r}$,
and send 1 qudit in it. Then the dealer is given a quantum secret
$\ket{\vec{s}}$ for $\vec{s} =(s) \in \mathbf{F}_4^1$.
Then we have
\[
g_{r,s}(x) = \frac{r-s}{\alpha_1} x + s.
\]
The unitary map $U_{\mathrm{enc}}$ sends
$\ket{r}\ket{s}$ to $\ket{\frac{r-s}{\alpha_1} \alpha_2 + s}
\ket{\frac{r-s}{\alpha_1} \alpha_3 + s}$.
Now it is clear that the left most share can be distributed before the
secret $\ket{s}$ is given, in contrast to Section \ref{ex:original}.

\section{Advance sharing with quantum ramp secret sharing scheme in \protect\cite{MATSUMOTO14STRONG}}\label{sec3}
\subsection{Zhang and Matsumoto's scheme}
As mentioned before,
Zhang and Matsumoto \cite{MATSUMOTO14STRONG} also proposed
a $(k,L,n)$ ramp quantum secret sharing scheme.
Their proposal has strong security in contrast to Ogawa et al.\ \cite{OGAWA05},
and the maximum possible number of participants was smaller,
that is, $n \leq q-L$.
As in Section \ref{sec2}, $\alpha_1$, \ldots, $\alpha_n$
are pairwise distinct elements in $\mathbf{F}_q$.
We choose extra elements $\beta_1$, \ldots, $\beta_L$
such that all $\alpha_1$, \ldots, $\alpha_n$ and $\beta_1$, \ldots, $\beta_L$
are pairwise distinct. In \cite{MATSUMOTO14STRONG}
any of $\alpha_1$, \ldots, $\alpha_n$ and $\beta_1$, \ldots, $\beta_L$ can
be $0 \in \mathbf{F}_q$.
For $\vec{s} \in \mathbf{F}_q^L$, we define
\begin{equation}
  D_{\mathrm{ZM}}(\vec{s}) = \{ \vec{c} \in \mathbf{F}_q^k : f_{\vec{c}}(\beta_i) = s_i
  \mbox{ for } i=1,\ldots, L\}, \label{eq100}
\end{equation}
where $f_{\vec{c}}(x)$ is as defined in Eq.\ (\ref{eq9}).
For a given quantum secret $\ket{\vec{s}}$,
in \cite{MATSUMOTO14STRONG} the quantum state of $n$ shares are obtained as
\begin{equation}
  \frac{1}{\sqrt{q^{k-L}}}\sum_{\vec{c}\in D_{\mathrm{ZM}}(\vec{s})} \ket{f_{\vec{c}}(\alpha_1)} \otimes \ket{f_{\vec{c}}(\alpha_2)} \otimes \cdots \otimes \ket{f_{\vec{c}}(\alpha_n)}, \label{eq101}
\end{equation}

\subsection{Example of the original encoding}\label{ex:original2}
Let $q=7$, $n=4$, $k=3$, $L=2$,
$(\alpha_1$, \ldots, $\alpha_4) =(6$, $2$, $4$, $5)$ and
$(\beta_1$, $\beta_2) = (1$, $3)$.
This example is the same as \cite[Example 2]{MATSUMOTO14STRONG}.
For $\vec{s} = (s_1$, $s_2)$,  
$D_{\mathrm{ZM}}(\vec{s})$ contains $(c_1$, \ldots, $c_3)$ if and only if
\[
\left\{
\begin{array}{rcl}
  c_1 + c_2 + c_3 &=& s_1,\\
  c_1 + 3c_2 + 2c_3&=& s_2.
\end{array}
\right.
\]
The solution of the above system of linear equations for $(c_1$, \ldots, $c_3)$ is
$D_{\mathrm{ZM}}(\vec{s}) = \{ (5s_1 + 3s_2 +3c_3,3s_1 + 4s_2+3c_3,c_3) : c_3 \in \mathbf{F}_7 \}$,
which is the same \cite[Eq.\ (3)]{MATSUMOTO14STRONG}.
For a given quantum state $\ket{\vec{s}} = \ket{s_1} \otimes \ket{s_2}$,
the quantum state of $4$ shares is (the normalizing constant $1/\sqrt{7}$ is omitted)
$\sum_{c_3 = 0}^6 \otimes_{i=1}^4 \ket{\alpha_i^2c_3  + \alpha_i (3s_1 + 4s_2+3c_3) +
  (5s_1 + 3s_2 +3c_3)}
= \ket{2s_1+6s_2} \otimes \ket{4s_1 +4s_2} \otimes  \ket{3s_1 + 5s_2} \otimes \ket{6s_1 + 2s_2} + \cdots $,
which is equivalent to \cite[Eq.\ (5)]{MATSUMOTO14STRONG}.
Since the quantum state of every qudit seems to depend on $s_1$ and $s_2$,
from the encoding procedure in \cite{MATSUMOTO14STRONG}, it seems unclear
how one share is advance shareable.

\subsection{Proposed advance sharing procedure for \protect\cite{MATSUMOTO14STRONG}}
We consider to distribute $k-L$ shares before $\ket{\vec{s}}$ is given to the
dealer. By reassigning indices, we may assume that the dealer
wants to advance share the 1st to the $(k-L)$-th shares.
Let $\ket{\Psi}$ as defined in Eq.\ (\ref{eq31}).

Suppose that the quantum secret is
$\ket{\vec{s}} = \ket{s_1} \otimes \cdots \otimes \ket{s_L} \in
\mathcal{H}_q^{\otimes L}$
with $\vec{s}=(s_1$, \ldots, $s_L) \in \mathbf{F}_q^L$.
For $\vec{r}\in \mathbf{F}_q^{k-L}$
and $\vec{s}$, let $h_{\vec{r},\vec{s}}(x)$ be a  univariate polynomial
  such that $h_{\vec{r},\vec{s}}(x) \in \{ f_{\vec{c}}(x)$ : $\vec{c}\in D_{\mathrm{ZM}}(\vec{s})\}$ and $h_{\vec{r},\vec{s}}(x)(\alpha_i) = r_i$
  for $i=1$, \ldots, $k-L$.

  \begin{proposition}\label{lem12}
    For given $\vec{r}$ and $\vec{s}$,
    the polynomial $h_{\vec{r},\vec{s}}(x)$ exists uniquely.
  \end{proposition}
  \begin{IEEEproof}
  Let $\mathcal{P}_k$ as defined in Lemma \ref{lem11}, and 
    consider a polynomial  $f(x) \in \mathcal{P}_k$.
    The condition $f(x) \in  \{ f_{\vec{c}}(x)$ : $\vec{c}\in D_{\mathrm{ZM}}(\vec{s})\}$
    means that $f(\beta_i) = s_i$ for $i=1$, \ldots, $L$.
Together with the conditions $f(\alpha_i) = r_i$
for $i=1$, \ldots, $k-L$, the required $h_{\vec{r},\vec{s}}(x)$
is a polynomial $f(x)\in \mathcal{P}_k$ such that $\mathrm{ev}'(f)
= (s_1$, \ldots, $s_L$, $r_1$, \ldots, $r_{k-L})$,
where $\mathrm{ev}'(f) = (f(\beta_1)$, \ldots, $f(\beta_{L})$,
$f(\alpha_1)$, \ldots, $f(\alpha_{k-L}))$.
By Lemma \ref{lem11}, we can see that such a polynomial $f(x)$
exists uniquely, which is $h_{\vec{r},\vec{s}}(x)$.
  \end{IEEEproof}

\begin{remark}\label{rem11}
Similarly to Remark \ref{rem10},
we will give a more explict description of $h_{\vec{r},\vec{s}}(x)$,
which is determined by the equality condition
$(h_{\vec{r},\vec{s}}(\beta_1)$, \ldots, $h_{\vec{r},\vec{s}}(\beta_L)$,
$h_{\vec{r},\vec{s}}(\alpha_1)$,
\ldots, $h_{\vec{r},\vec{s}}(\alpha_{k-L}))= (s_1$, \ldots, $s_L$, $r_1$, \ldots, $r_{k-L})$.
The determination of $h_{\vec{r},\vec{s}}(x)$
is known as the Lagrange interpolation \cite[Chapter 13]{stinson06},
which can be computed as follows.
In order to use the Lagrange interpolation formula,
define $\beta_{L+j} = \alpha_j$ for $j=1$, \ldots, $k-L$, and
\[
\ell_i(x) = \prod_{1 \leq j \leq k, j\neq i} \frac{x-\beta_j}{\beta_i - \beta_j}.
\]
Now $h_{\vec{r},\vec{s}}(x)$ can be obtained as
\[
h_{\vec{r},\vec{s}}(x)= \sum_{i=1}^{L} s_i \ell_i(x) + \sum_{i=1}^{k-L} r_i \ell_{i+L}(x).
\]
\end{remark}

Let us describe a new encoding procedure enabling advance sharing.
    By Proposition \ref{lem12},
    we can define a unitary map $U_{\mathrm{ZM,enc}}$ from
    $\mathcal{H}_q^{\otimes k}$ to $\mathcal{H}_q^{\otimes (n-k+L)}$ 
    sending a quantum state $\ket{\vec{r}}\ket{\vec{s}}$
    to $\ket{h_{\vec{r},\vec{s}}(\alpha_{k-L+1})} \otimes \cdots \otimes
    \ket{h_{\vec{r},\vec{s}}(\alpha_n)}$.
    Note that $k=n-k+L$ by \cite[Lemma 2]{OGAWA05}.
  
The proposed procedure for advance sharing is as follows:
\begin{enumerate}
\item The dealer distribute $k-L$ qudits in the left half of Eq.\ (\ref{eq31}).
\item After the quantum secret $\ket{\vec{s}}$ is given,
the dealer applies $U_{\mathrm{ZM,enc}}$ to the remaining half of Eq.\ (\ref{eq31})
and $\ket{\vec{s}}$. Then the quantum state of all the $n$ shares becomes
  \begin{equation}
  \frac{1}{\sqrt{q^{k-L}}}
  \sum_{\vec{r}\in \mathbf{F}_q^{k-L}} \ket{\vec{r}}\otimes \ket{h_{\vec{r},\vec{s}}(\alpha_{k-L+1})}
    \otimes \cdots \otimes \ket{h_{\vec{r},\vec{s}}(\alpha_{n})}. \label{eq143}
  \end{equation}
\end{enumerate}

\begin{remark}\label{rem2}
When $k-L-1$ or a fewer shares are advance shared,
the dealer can simply keep some shares in the advance sharing phase in our
proposal as in Remark \ref{rem1}.
On the other hand, as in Remark \ref{rem1},
it is impossible to advance share $k-L+1$ or more shares,
because in Zhang and Matsumoto's $(k,L,n)$ scheme
\cite{MATSUMOTO14STRONG}, any $k-L+1$ or more shares
cannot form a forbidden set.
Therefore, similarly to Remark \ref{rem1},
our proposal makes advance shareable sets as large as possible.
\end{remark}

\subsection{Proof of correctness}\label{sec35}
By the same reason explained at the beginning of Section \ref{sec25},
in this subsection, we will prove that
the proposed procedure gives the same quantum state of shares
as the original \cite{MATSUMOTO14STRONG}
for a given quantum secret $\ket{\vec{s}}$.
In Eq.\ (\ref{eq101}), the set of indices appearing in the quantum state is
\begin{equation}
\{(f_{\vec{c}}(\alpha_1), \ldots, f_{\vec{c}}(\alpha_n) ) : \vec{c}\in D_{\mathrm{ZM}}(\vec{s})\}.
\label{eq141}
\end{equation}
  In Eq.\ (\ref{eq143}), the set of indices appearing in the quantum state is
\begin{equation}
\{(r_1, \ldots, r_{k-L},   h_{\vec{r},\vec{s}}(\alpha_{k-L+1}), \ldots, h_{\vec{r},\vec{s}}(\alpha_{n}) : \vec{r} \in \mathbf{F}_q^{k-L} \}. \label{eq152}
\end{equation}
\begin{theorem}
For a fixed quantum secret $\ket{\vec{s}}$ for $\vec{s} \in \mathbf{F}_q^L$,
the proposed procedure gives the same quantum state of shares
as the original \cite{MATSUMOTO14STRONG}.
\end{theorem}
\begin{IEEEproof}
In order to prove the theorem,
it is enough to show that the sets (\ref{eq141}) and (\ref{eq152}) are the same.
For fixed $\vec{s} \in \mathbf{F}_q^L$ and
$\vec{r} \in \mathbf{F}_q^{k-L}$, by Proposition \ref{lem12}
there exists a unique polynomial $f_{\vec{c}}(x)$ with $\vec{c} \in D_{\mathrm{ZM}}(\vec{s}) $.
This correspondence gives a bijection between $\vec{r} \in \mathbf{F}_q^{k-L}$ to
$f_{\vec{c}}(x)$ with $\vec{c} \in D(\vec{s}) $.
By the definition of $h_{\vec{r},\vec{s}}$, we have
\begin{eqnarray*}
&&(r_1, \ldots, r_{k-L},   h_{\vec{r},\vec{s}}(\alpha_{k-L+1}), \ldots, h_{\vec{r},\vec{s}}(\alpha_{n}))\\
&=& (h_{\vec{r},\vec{s}}(\alpha_1), \ldots, h_{\vec{r},\vec{s}}(\alpha_{n}))\\
&=&(f_{\vec{c}}(\alpha_1), \ldots, f_{\vec{c}}(\alpha_{n})),
\end{eqnarray*}
which shows the theorem, where $\vec{c}$ is chosen according to $\vec{r}$.
\end{IEEEproof}

\subsection{Example of the proposed encoding}
We reuse definitions from Section \ref{ex:original2}.
Let $q=7$, $n=4$, $k=3$, $L=2$,
$(\alpha_1$, \ldots, $\alpha_4) =(6$, $2$, $4$, $5)$ and
$(\beta_1$, $\beta_2) = (1$, $3)$.
In our proposal,
the dealer prepares 2-qudit entangled state
$\frac{1}{\sqrt{7}}\sum_{r \in \mathbf{F}_7} \ket{r}\ket{r}$,
and send 1 qudit in it. Then the dealer is given a quantum secret
$\ket{\vec{s}}$ for $\vec{s} =(s_1,s_2) \in \mathbf{F}_7^2$.
Then we have
\[
h_{r,s_1,s_2}(x) = (r-2s_1+s_2)x^2 + (4s_1-4r)x+(3r-s_1-s_2),
\]
because $h_{r,s_1,s_2}(x)$ satisfies
$h_{r,s_1,s_2}(\alpha_1)=h_{r,s_1,s_2}(6)=r$ , $h_{r,s_1,s_2}(\beta_1)=h_{r,s_1,s_2}(1)=s_1$,
and $h_{r,s_1,s_2}(\beta_2)=h_{r,s_1,s_2}(3)=s_2$.
The unitary map $U_{\mathrm{ZM,enc}}$ sends
$\ket{r}\ket{s_1}\ket{s_2}$ to $\ket{h_{r,s_1,s_2}(\alpha_2)}
\ket{h_{r,s_1,s_2}(\alpha_3)}\ket{h_{r,s_1,s_2}(\alpha_3)}$.
Now it is clear that the left most share can be distributed before the
secret $\ket{\vec{s}}$ is given, in contrast to Section \ref{ex:original2}.

\section{Concluding remarks}\label{sec4}
In this paper, we proposed new encoding procedures  for Ogawa et al.'s \cite{OGAWA05}
and Zhang-Matsumoto \cite{MATSUMOTO14STRONG} ramp
quantum secret sharing schemes, and allowed the dealer to distribute shares
in those schemes before secrets are given.
This enhances applicability of the secret sharing schemes by Ogawa et al.  \cite{OGAWA05}
and by Zhang and Matsumoto \cite{MATSUMOTO14STRONG}
to wider scenarios in which some participants are unavailable when the dealer obtains a secret.
We also proved that the new proposed encoding procedures
retain desirable properties, such as coding rates, access structures, strong security
from the original schemes.
We also proved that the advance shareable sets are made as large as
possible in the proposed encoding procedures.

Similarly to \cite{shibata24},
there could exist a general method to construct advance sharing
procedure for any stabilizer-based quantum secret sharing,
while retaining the correspondence between secrets and shares.
Its investigation could be a future research agenda.

As mentioned in Section \ref{sec21},
it seems also important to study
advance sharing with other scenarios of quantum secret sharing,
for example, \cite{crepeau02,benor06,crepeau05}.
It is also another research agenda.

\section*{Acknowledgment}
The author would like to deeply thank reviewers' helpful comments
that improved this paper.


\begin{thebibliography}{10}
\providecommand{\url}[1]{#1}
\csname url@samestyle\endcsname
\providecommand{\newblock}{\relax}
\providecommand{\bibinfo}[2]{#2}
\providecommand{\BIBentrySTDinterwordspacing}{\spaceskip=0pt\relax}
\providecommand{\BIBentryALTinterwordstretchfactor}{4}
\providecommand{\BIBentryALTinterwordspacing}{\spaceskip=\fontdimen2\font plus
\BIBentryALTinterwordstretchfactor\fontdimen3\font minus
  \fontdimen4\font\relax}
\providecommand{\BIBforeignlanguage}[2]{{%
\expandafter\ifx\csname l@#1\endcsname\relax
\typeout{** WARNING: IEEEtran.bst: No hyphenation pattern has been}%
\typeout{** loaded for the language `#1'. Using the pattern for}%
\typeout{** the default language instead.}%
\else
\language=\csname l@#1\endcsname
\fi
#2}}
\providecommand{\BIBdecl}{\relax}
\BIBdecl

\bibitem{googleqc}
F.~Arute \emph{et~al.}, ``Quantum supremacy using a programmable
  superconducting processor,'' \emph{Nature}, vol. 574, pp. 505--510, Oct.
  2019. doi:10.1038/s41586-019-1666-5

\bibitem{attasena2017}
V.~Attasena, J.~Darmont, and N.~Harbi, ``Secret sharing for cloud data
  security: {A} survey,'' \emph{The VLDB Journal}, vol.~26, no.~5, pp.
  657--681, Oct. 2017. doi:10.1007/s00778-017-0470-9

\bibitem{cleve99}
R.~Cleve, D.~Gottesman, and H.-K. Lo, ``How to share a quantum secret,''
  \emph{Phys. Rev. Lett.}, vol.~83, no.~3, pp. 648--651, Jul. 1999.
  doi:10.1103/PhysRevLett.83.648

\bibitem{gottesman00}
D.~Gottesman, ``Theory of quantum secret sharing,'' \emph{Phys. Rev. A},
  vol.~61, no.~4,  042311, Mar. 2000. doi:10.1103/PhysRevA.61.042311

\bibitem{shamir79}
A.~Shamir, ``How to share a secret,'' \emph{Comm. ACM}, vol.~22, no.~11, pp.
  612--613, Nov. 1979. doi:10.1145/359168.359176

\bibitem{blakley79}
G.~R. Blakley, ``Safeguarding cryptographic keys,'' in \emph{AFIPS
  International Workshop on Managing Requirements Knowledge}.\hskip 1em plus
  0.5em minus 0.4em\relax IEEE, Jun. 1979, pp. 242--269.
  doi:10.1109/MARK.1979.8817296

\bibitem{OGAWA05}
T.~Ogawa, A.~Sasaki, M.~Iwamoto, and H.~Yamamoto, ``Quantum secret sharing
  schemes and reversibility of quantum operations,'' \emph{Phys. Rev. A},
  vol.~72, no.~3, 032318, Sep. 2005.
  doi:10.1103/PhysRevA.72.032318

\bibitem{iwamoto06}
M.~Iwamoto and H.~Yamamoto, ``Strongly secure ramp secret sharing schemes for
  general access structures,'' \emph{Inform. Process. Lett.}, vol.~97, no.~2,
  pp. 52--57, Jan. 2006. doi:10.1016/j.ipl.2005.09.012

\bibitem{yamamoto86}
H.~Yamamoto, ``Secret sharing system using $(k, l, n)$ threshold scheme,''
  \emph{Electronics and Communications in Japan (Part I: Communications)},
  vol.~69, no.~9, pp. 313--317, 1985.
  doi:10.1002/ecja.4410690906

\bibitem{MATSUMOTO14STRONG}
P.~Zhang and R.~Matsumoto, ``Quantum strongly secure ramp secret sharing,''
  \emph{Quantum Information Processing}, vol.~14, no.~2, pp. 715--729, Feb.
  2015. doi:10.1007/s11128-014-0863-2

\bibitem{miyajima22}
R.~Miyajima and R.~Matsumoto, ``Advance sharing of quantum shares for classical
  secrets,'' \emph{IEEE Acess}, 2022, 10.1109/ACCESS.2022.3204389.

\bibitem{lie20}
S.~H. Lie and H.~Jeong, ``Randomness cost of masking quantum information and
  the information conservation law,'' \emph{Phys. Rev. A}, vol. 101, 052322,
  May 2020. doi:10.1103/PhysRevA.101.052322

\bibitem{shibata24}
M.~Shibata and R.~Matsumoto, ``Advance sharing of quantum shares for quantum
  secrets,'' \emph{IEICE Trans. Fundamentals.}, vol. E107-A, no. 8,
  pp.\ 1247--1254, Aug. 2024.
  doi:10.1587/transfun.2023EAP1041

\bibitem{isita2024}
S.~Masumori and R.~Matsumoto, ``Advance sharing with {Ogawa} et al.'s ramp
  quantum secret sharing scheme for prime-dimensional quantum systems,'' in
  \emph{Proc.\ 2024 International Symposium on Information Theory and its
  Applications}, National Taiwan University of Science and Technology, Taiwan,
  Nov. 2024, pp. 50--52.

\bibitem{stinson06}
D.~R. Stinson, \emph{Cryptography Theory and Practice}, 3rd~ed.\hskip 1em plus
  0.5em minus 0.4em\relax Chapman \& Hall/CRC, 2006.
doi:10.1201/9781420057133

\bibitem{hillery99}
M.~Hillery, V.~Bu\v{z}ek, and A.~Berthiaume, ``Quantum secret sharing,''
  \emph{Phys. Rev. A}, vol.~59, pp. 1829--1834, Mar. 1999.
  doi:10.1103/PhysRevA.59.1829

\bibitem{qsdcsurvey2024}
D.~Pan, G.-L. Long, L.~Yin, Y.-B. Sheng, D.~Ruan, S.~X. Ng, J.~Lu, and
  L.~Hanzo, ``The evolution of quantum secure direct communication: On the road
  to the qinternet,'' \emph{IEEE Communications Surveys \& Tutorials}, vol.~26,
  no.~3, pp. 1898--1949, 2024.
  doi:10.1109/COMST.2024.3367535

\bibitem{crepeau02}
C.~Cr\'epeau, D.~Gottesman, and A.~Smith, ``Secure multi-party quantum
  computation,'' in \emph{Proceedings of the Thiry-Fourth Annual ACM Symposium
  on Theory of Computing}.\hskip 1em plus 0.5em minus 0.4em\relax New York, NY,
  USA: Association for Computing Machinery, 2002, p. 643–652.
  doi:10.1145/509907.510000

\bibitem{benor06}
M.~Ben-Or, C.~Cr\'epeau, D.~Gottesman, A.~Hassidim, and A.~Smith, ``Secure
  multiparty quantum computation with (only) a strict honest majority,'' in
  \emph{2006 47th Annual IEEE Symposium on Foundations of Computer Science
  (FOCS'06)}, 2006, pp. 249--260. doi:10.1109/FOCS.2006.68

\bibitem{crepeau05}
C.~Cr\'epeau, D.~Gottesman, and A.~Smith, ``Approximate quantum
  error-correcting codes and secret sharing schemes,'' in \emph{Advances in
  Cryptology -- EUROCRYPT 2005}, R.~Cramer, Ed.\hskip 1em plus 0.5em minus
  0.4em\relax Berlin, Heidelberg: Springer Berlin Heidelberg, 2005, pp.
  285--301.
  doi:10.1007/11426639\_17

\end{thebibliography}


\begin{IEEEbiography}[{\includegraphics[width=1in,height=1.25in,clip,keepaspectratio]{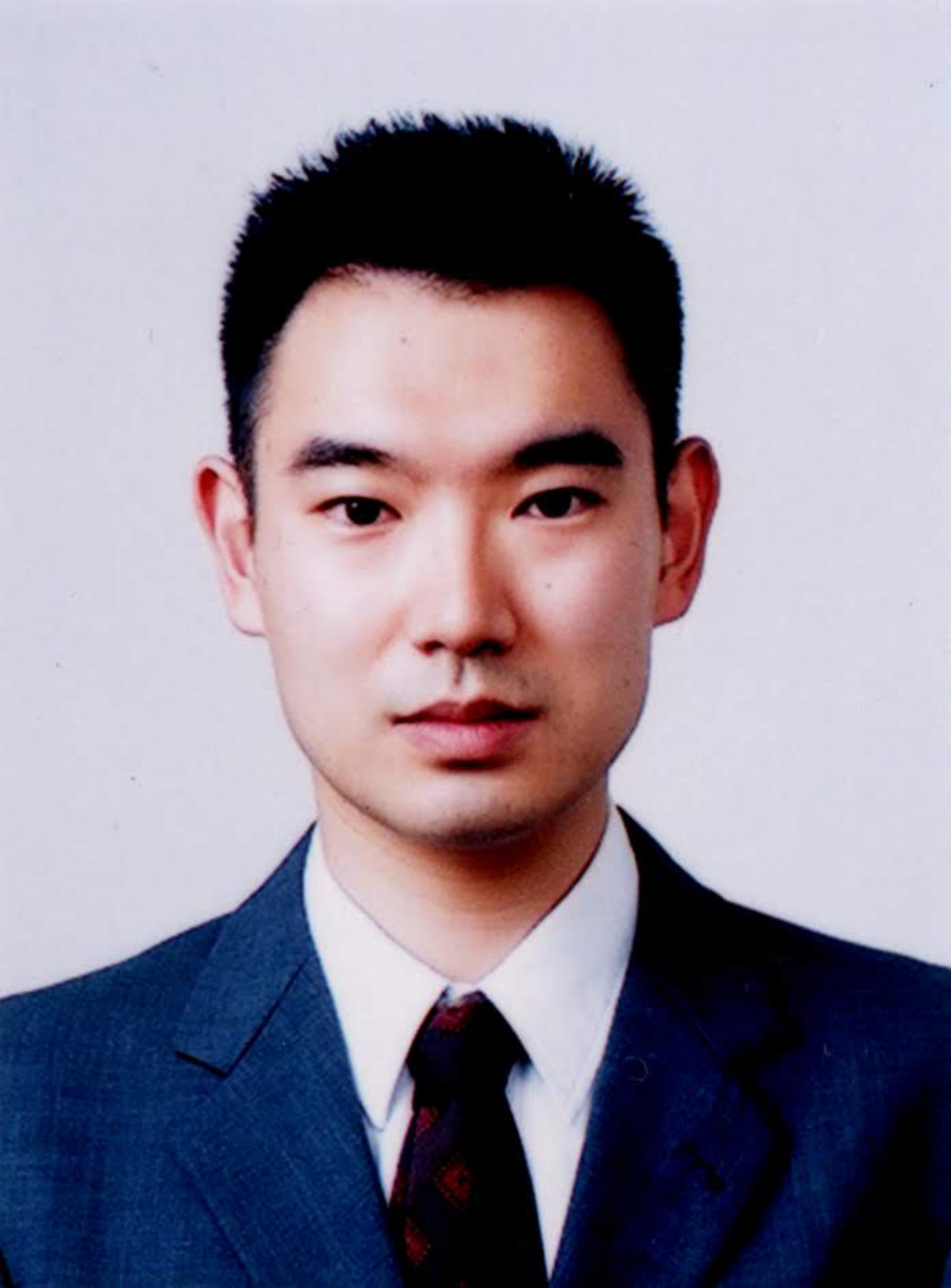}}]{Ryutaroh Matsumoto} (Member, IEEE)
  was born in Nagoya, Japan, in November 1973.
He received the B.E.\ degree in computer science,
the M.E.\ degree in information processing, and the
Ph.D.\ degree in electrical and electronic engineering from the Tokyo Institute of Technology, Tokyo,
Japan, in 1996, 1998, and 2001, respectively.

From 2001 to 2004, he was an Assistant Professor with the Department of Information and
Communications Engineering, Tokyo Institute of
Technology, where he was an Associate Professor, from 2004 to 2017.
From 2017 to 2020, he was an Associate Professor with the Department of
Information and Communication Engineering, Nagoya University, Nagoya,
Japan. Since 2020, he has been an Associate Professor, and promoted to
a Full Professor, in August 2022, with the Department of Information and
Communications Engineering, Tokyo Institute of Technology. In 2011 and
2014, he was as a Velux Visiting Professor with the Department of Mathematical Sciences, Aalborg University, Aalborg, Denmark. His research interests
include error-correcting codes, quantum information theory, information
theoretic security, and communication theory.

Dr.\ Matsumoto was a recipient of the Young Engineer Award from IEICE
and the Ericsson Young Scientist Award from Ericsson Japan in 2001. He was
also a recipient of the Best Paper Awards from IEICE in 2001, 2008, 2011,
and 2014.
\end{IEEEbiography}

\EOD

\end{document}